\documentclass[preprint,showpacs,preprintnumbers,amsmath,amssymb, showkeys,superscriptaddress]{revtex4}
\usepackage{graphicx}
\usepackage{dcolumn}
\usepackage{bm}
\usepackage{hyperref}

\begin{document}

\title{The Hoyle state in relativistic dissociation of light nuclei \footnote[1]{Submitted to Physics of Atomic Nuclei}}

\author{A.A. Zaitsev}
\affiliation{Joint Institute for Nuclear Research (JINR), Dubna, Russia}
\affiliation{Lebedev Physical Institute, Russian Academy of science, Moscow, Russia}
\author{P.I. Zarubin}
\email{zarubin@lhe.jinr.ru}
\affiliation{Joint Institute for Nuclear Research (JINR), Dubna, Russia}
\affiliation{Lebedev Physical Institute, Russian Academy of science, Moscow, Russia}

\begin{abstract}
	In the context of the search for triples of relativistic $\alpha$-particles in the Hoyle state, the analysis of available data on the dissociation of the nuclei ${}^{12}$C, ${}^{16}$O and ${}^{22}$Ne in the nuclear emulsion was carried out. The Hoyle state is identified by the invariant mass calculated from pair angles of expansion in $\alpha$-triples in the approximation of the conservation of the momentum per nucleon of the parent nucleus. The contribution of the Hoyle state to the dissociation of ${}^{12}$C $\to$ 3$\alpha$ is 11\%. In the case of the coherent dissociation of ${}^{16}$O $\to$ 4$\alpha$ it reaches 22\% when the portion of the channel ${}^{16}$O $\to$ 2${}^{8}$Be is equal to 5\%.
\end{abstract}

\maketitle 

\section{Introduction}
Irradiations of stacks of nuclear track emulsion (NTE) in beams of light relativistic nuclei were performed in the 70-80s at the JINR Synchrophasotron and Bevalac (LBL, USA). The use of NTE to study the interactions of the gold and lead nuclei continued in the 80-90s at the AGS (BNL, USA) and SPS (CERN) accelerators. Observations in NTE of tracks of charged particles in a full solid angle and practically without a threshold made it possible to determine the contours of a complex picture of the collision of relativistic nuclei. Special attention was paid to central nuclear collisions. The subsequent development of this area on the basis of large-scale electronic experiments is widely known. At the same time, the results obtained by the NTE method as well as the irradiated layers themselves and the files with measurement results retain their uniqueness with respect to the structure of nuclear fragmentation.

In peripheral interactions of nuclei, in which the charge of the incident nucleus is distributed between its fragments, the individual features of the incident nuclei are reflected. They are observed in NTE as often and completely as central collisions, so there is a fundamental possibility in the cone of relativistic fragmentation to study the nuclear structure. However, in this aspect, the use of traditional spectrometers was extremely limited. The difficulties encountered are of fundamental nature. They are caused by a dramatic decrease in the ionization of relativistic fragments in an extremely narrow fragmentation cone, and, often, by an approximate coincidence in the magnetic rigidity of fragments and beam nuclei. For these reasons, measurements were carried out with the registration of single relativistic fragments with charges close to the charge of the studied nucleus.

The pause in obtaining data on the ''fine`` structure of relativistic fragmentation motivated further NTE irradiations in beams of light nuclei of the JINR Nuclotron including radioactive ones. Since the early 2000s, the BECQUEREL experiment aimed at systematic study of peripheral interactions of relativistic nuclei by the NTE method has been started. The analysis of peripheral interactions in longitudinally irradiated NTE layers allowed one to study cluster features of a whole family of light nuclei, including neutron-deficient ones, in a single approach (reviews \cite{1,2}). The possibility of analyzing such ensembles is a prerequisite for testing the concepts developed in nuclear physics and nuclear astrophysics. The role of unstable cores ${}^{8}$Be and ${}^{9}$B in their structure was established. In the dissociation of the ${}^{10}$C nucleus, an indication of the ${}^{9}$B$p$ resonance at energy of about 4 MeV was found.

The decisive factor for reconstructing the decays of ${}^{8}$Be and ${}^{9}$B nuclei among fragments of a relativistic projectile nucleus is the best spatial resolution (about 0.5 $\mu$m) provided by the NTE technique. Decays are identified by the invariant mass $M^*$, determined by the sum of all products of 4-momenta $P_i$ of relativistic fragments He and H. Subtraction of the sum of masses of fragments $Q = M^* - M$ is a matter of convenience. The components $P_i$ of are determined from the angles of emission of He and H fragments assuming that they maintain momentum per nucleon of the projectile (or its velocity). Then the invariant mass of the considered ensemble of fragments is determined by the angles of their expansion. ${}^{9}$B $\to$ ${}^{8}$Be$p$ decays can be considered as a pure source of ${}^{8}$Be nuclei. Their analysis allowed us to confirm the criterion $Q_{2\alpha}$ $<$ 0.2 MeV for the selection of ${}^{8}$Be which takes into account the adopted approximation and resolution of the method \cite{3}.

The successful reconstruction of the ${}^{8}$Be and ${}^{9}$B decays allows one to take the next step — to search in relativistic dissociation ${}^{12}$C $\to$ 3$\alpha$ for triples of $\alpha$-particles in the Hoyle state (HS). This state is the second and first unbound excitation 0$^+_2$ of the ${}^{12}$C nucleus. The significance of this short-lived state of three real $\alpha$-particles and the status of its research are presented in the review \cite{4}. The HS features such as isolation in the initial part of the ${}^{12}$C excitation spectrum, lowest decay energy and its narrow width (378 keV and 8.5 eV) indicate its similarity with the 2$\alpha$-particle nucleus ${}^{8}$Be (91 keV and 5.6 eV). ${}^{8}$Be is an indispensable product of HS decays. It can be assumed that HS is not limited to ${}^{12}$C excitation but it can also appear as a 3$\alpha$-partial analog of ${}^{8}$Be in relativistic fragmentation of heavier nuclei.

Interest in HS is motivated by the concept of $\alpha$-partial Bose-Einstein condensate (review \cite{5}), the status of which is presented in \cite{6}. As the simplest forms of such a condensate the ground state of the unstable ${}^{8}$Be nucleus and, after it, HS are suggested. Continuing the ${}^{8}$Be and HS branches, it is assumed that the condensate 4$\alpha$ state is the 6th excited state 0$^+_6$ of the ${}^{16}$O nucleus, located 700 keV above the 4$\alpha$ threshold. Then, the condensate decomposition could go in the sequence ${}^{16}$O(0$^+_6$) $\to$ ${}^{12}$C(0$^+_2$) $\to$ ${}^{8}$Be(0$^+_2$) $\to$ 2$\alpha$.

The fact of HS generation may reflect both the presence of three weakly bound $\alpha$-particles in the 0S-state in the parent nucleus as well as arise through the excited fragment ${}^{12}$C$^*$($\to$3$\alpha$) or be a product of the interaction of $\alpha$-particles in the final state. These options require theoretical consideration. Experimentally, the general question is as follows. Can the fragmentation of relativistic nuclei serve as a ``factory'' for the generation of ensembles of $\alpha$-particles of increasing multiplicity at the lower limit of nuclear temperature? Further, in the context of the HS problem, distributions of invariant mass $Q$(2-4)$\alpha$ of $\alpha$-partial pairs, triples and quartets born in the dissociation of nuclei ${}^{12}$C, ${}^{16}$O and ${}^{22}$Ne will be presented.

\section{DISSOCIATION OF ${}^{12}$C NUCLEI}

For the ${}^{12}$C nucleus at an energy of 3.65 $A$ GeV there are measurements of emission angles of $\alpha$-particles made in the groups of G. M. Chernov (Tashkent) \cite{7} at 72 and A. Sh. Gaitinov ( Alma-Ata) in 114 events coherent dissociation of ${}^{12}$C $\to$ 3$\alpha$, not accompanied by fragments of target nuclei or generated mesons, which are briefly referred to as ``white'' stars. The search for such events was carried out in an accelerated manner along transverse strips of NTE layers. Thus, the contribution of ${}^{8}$Be $\to$ 2$\alpha$ decays by the smallest angles of scattering of $\alpha$-particles was determined \cite{7}. Figure 1 shows the distribution over the invariant mass of $\alpha$-pairs $Q_{2\alpha}$. In the $Q_{2\alpha}$ $<$ 0.2 MeV region, the contribution of the ${}^{8}$Be decays is 17 $\pm$ 1\%.

Recently, in collaboration with the group of N. G. Peresadko (FIAN), data on 238 3$\alpha$-stars, including 130 ``white'', have been added. In addition, there are NTE layers irradiated in the ${}^{12}$C  beam of the booster of the Institute of High Energy Physics (Protvino) at 420 $A$ MeV, which allow using an approach based on a variable invariant mass \cite{5}. In the latter case, emission angles are measured in 86 3$\alpha$-events found, including 36 ``white'' stars.

\begin{figure}
	\centerline{\includegraphics*[width=0.6\linewidth]{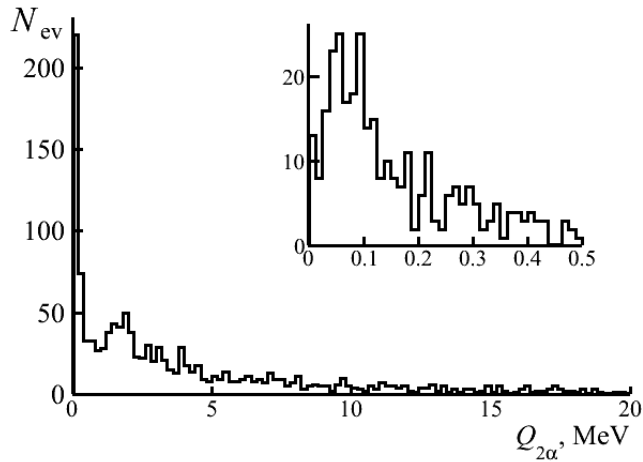}}
	\caption{Distribution over invariant mass $Q_{2\alpha}$ of $\alpha$-pairs in coherent dissociation ${}^{12}$C $\to$ 3$\alpha$ at 3.65 $A$ GeV (hatched); in the inset, part of the distribution $Q_{2\alpha}$ $<$ 0.5 MeV.}
\end{figure}

The distribution $Q_{3\alpha}$ for all 510 stars is shown in Fig. 2. The region $Q_{3\alpha}$ $<$ 10 MeV covering the ${}^{12}$C $\alpha$-particle excitations below the nucleon separation thresholds is described by the Rayleigh distribution with the parameter $\sigma_{Q(3\alpha)}$ = (3.9 $\pm$ 0.4) MeV. There is a peak in the region $Q_{3\alpha}$ $<$ 1 MeV (51 stars) where the HS signal is expected. For events at 3.65 $A$ GeV contributed to this peak the average value $\langle Q_{3\alpha} \rangle$ (RMS) is 397 $\pm$ 26 (166) keV, and at 420 $A$ MeV, 346 $\pm$ 28 (85) keV, respectively. According to the condition $Q_{3\alpha}$ $<$ 0.7 MeV 42 of 424 events at 3.65 $A$ GeV can be attributed to HS decays, and 9 at 420 $A$ MeV  (out of 86) including 5 ``white'' stars (out of 36). As a result, the contribution of HS decays to ${}^{12}$C $\to$ 3$\alpha$ dissociation is 10 $\pm$ 2\%.

\begin{figure}
	\centerline{\includegraphics*[width=0.6\linewidth]{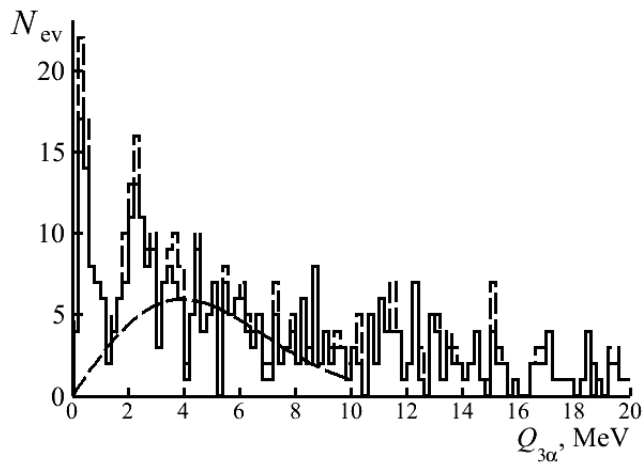}}
	\caption{Distribution over invariant mass $Q_{3\alpha}$ of $\alpha$-triples in the dissociation ${}^{12}$C $\to$ 3$\alpha$ at 3.65 $A$ GeV (shaded) and 420 $A$ MeV (added by a dotted line); line - Rayleigh distribution.}
\end{figure}

\section{COHERENT DISSOCIATION OF ${}^{16}$O NUCLEI}

The distribution $Q_{2\alpha}$ for all 2$\alpha$ combinations in 641 ``white'' star ${}^{16}$O $\to$ 4$\alpha$ according to the data \cite{8} is presented in Fig. 3. As in the case ${}^{12}$C $\to$ 3$\alpha$, for $Q_{2\alpha}$ $<$ 0.2 MeV there is a contribution of ${}^{8}$Be decays which manifests itself in 15 $\pm$ 1\% of events.

\begin{figure}
	\centerline{\includegraphics*[width=0.6\linewidth]{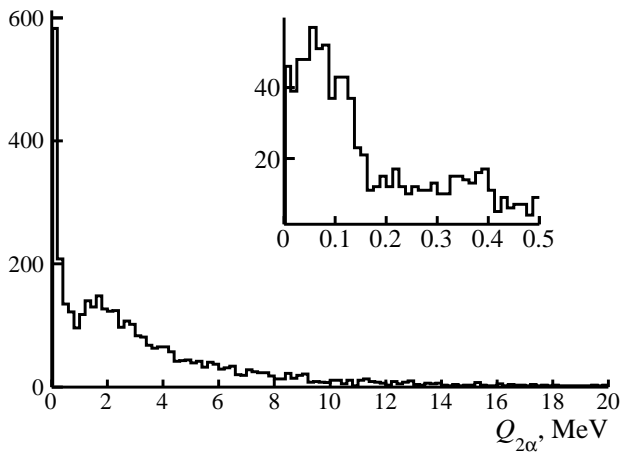}}
	\caption{Distribution of all $\alpha$-pairs in ``white'' stars ${}^{16}$O $\to$ 4$\alpha$ at 3.65 $A$ GeV over the invariant mass $Q_{2\alpha}$; in the inset, part of the distribution  $Q_{2\alpha}$ $<$ 0.5 MeV.}
\end{figure}

HS decays can manifest themselves in the dissociation ${}^{16}$O $\to$ ${}^{12}$C$^*$ ($\to$ 3$\alpha$) + $\alpha$. Figure 4 shows the $Q_{3\alpha}$ distribution of all 3$\alpha$ combinations. As in the ${}^{12}$C case, its main part with $Q_{3\alpha}$ $<$ 10 MeV is described by the Rayleigh distribution with the parameter $\sigma_{(Q3\alpha)}$ = (3.8 $\pm$ 0.2) MeV. It also has a peak at $Q_{3\alpha}$ $<$ 700 keV. The condition $Q_{2\alpha}$ $<$ 200 keV meaning at least one ${}^{8}$Be decay in a 4$\alpha$ event does not affect the statistics in this $Q_{3\alpha}$ range. The contribution to the peak of the combinatorial background estimated at 8\% is excluded. The remaining 139 events have an average value of $\langle Q_{3\alpha} \rangle$ = (349 $\pm$ 14) keV corresponding to HS and RMS 174 keV. In 9 events of them more than one 3$\alpha$-combination corresponds to the condition $Q_{3\alpha}$ $<$ 700 keV. In sum, the contribution of HS decays to the coherent dissociation of ${}^{16}$O $\to$ 4$\alpha$ is 22 $\pm$ 2\%. The distribution of $\alpha$-triples of HS over the total transverse momentum $P$T(HS) (Fig. 5) is described by the Rayleigh distribution with the parameter $\sigma_{PT}$(HS) = (191 $\pm$ 8) MeV/$c$ the value of which is characteristic for nuclear diffraction.

\begin{figure}
	\centerline{\includegraphics*[width=0.6\linewidth]{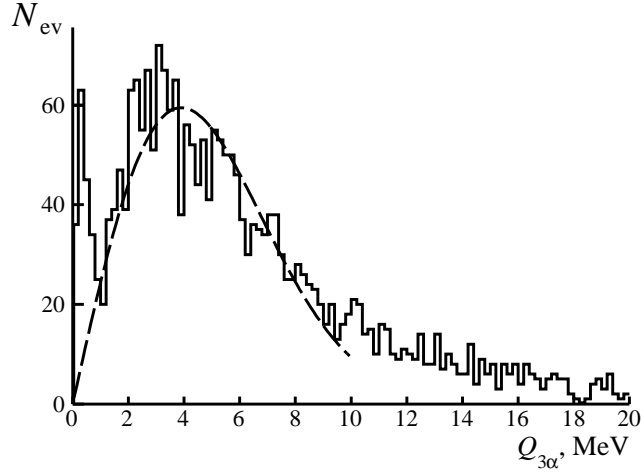}}
	\caption{Distribution of all $\alpha$-triples in ``white'' stars ${}^{16}$O $\to$ 4$\alpha$ at 3.65 $A$ GeV over invariant mass $Q_{3\alpha}$; line - Rayleigh distribution.}
\end{figure}

\begin{figure}
	\centerline{\includegraphics*[width=0.6\linewidth]{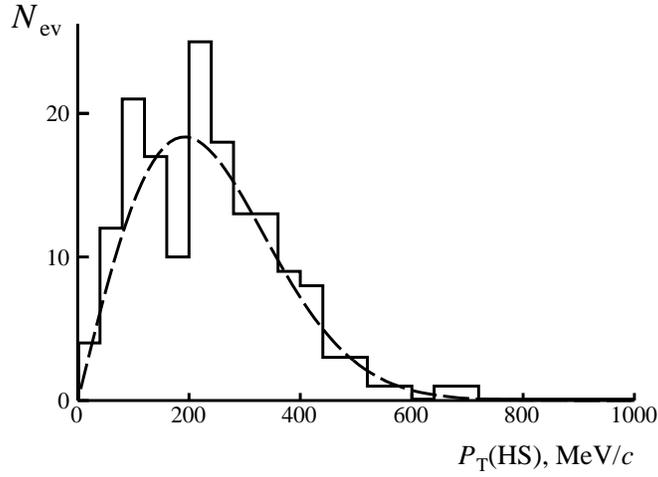}}
	\caption{Distribution of $\alpha$-triples $Q_{3\alpha}$ $<$ 0.7 MeV (HS) in ``white'' stars ${}^{16}$O $\to$ 4$\alpha$ at 3.65 $A$ GeV over total transverse momentum $P$T (HS); line - Rayleigh distribution.}
\end{figure}

\begin{figure}
	\centerline{\includegraphics*[width=0.6\linewidth]{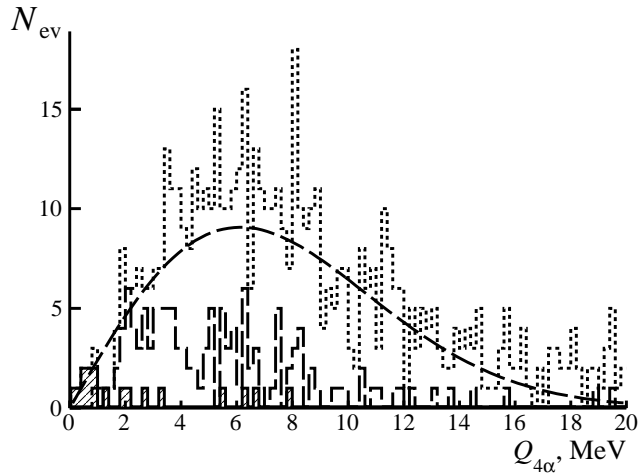}}
	\caption{Distribution over invariant mass $Q_{4\alpha}$ of ``white'' stars ${}^{16}$O $\to$ 4$\alpha$ at 3.65 $A$ GeV of all 4$\alpha$-quartets (points), $\alpha$HS events (dotted lines) and $\alpha$HS events $\varepsilon$ ($\alpha$HS) $<$ 45$^\circ$ (hatched); line - Rayleigh distribution.}
	\end{figure}

\begin{figure}
	\centerline{\includegraphics*[width=0.6\linewidth]{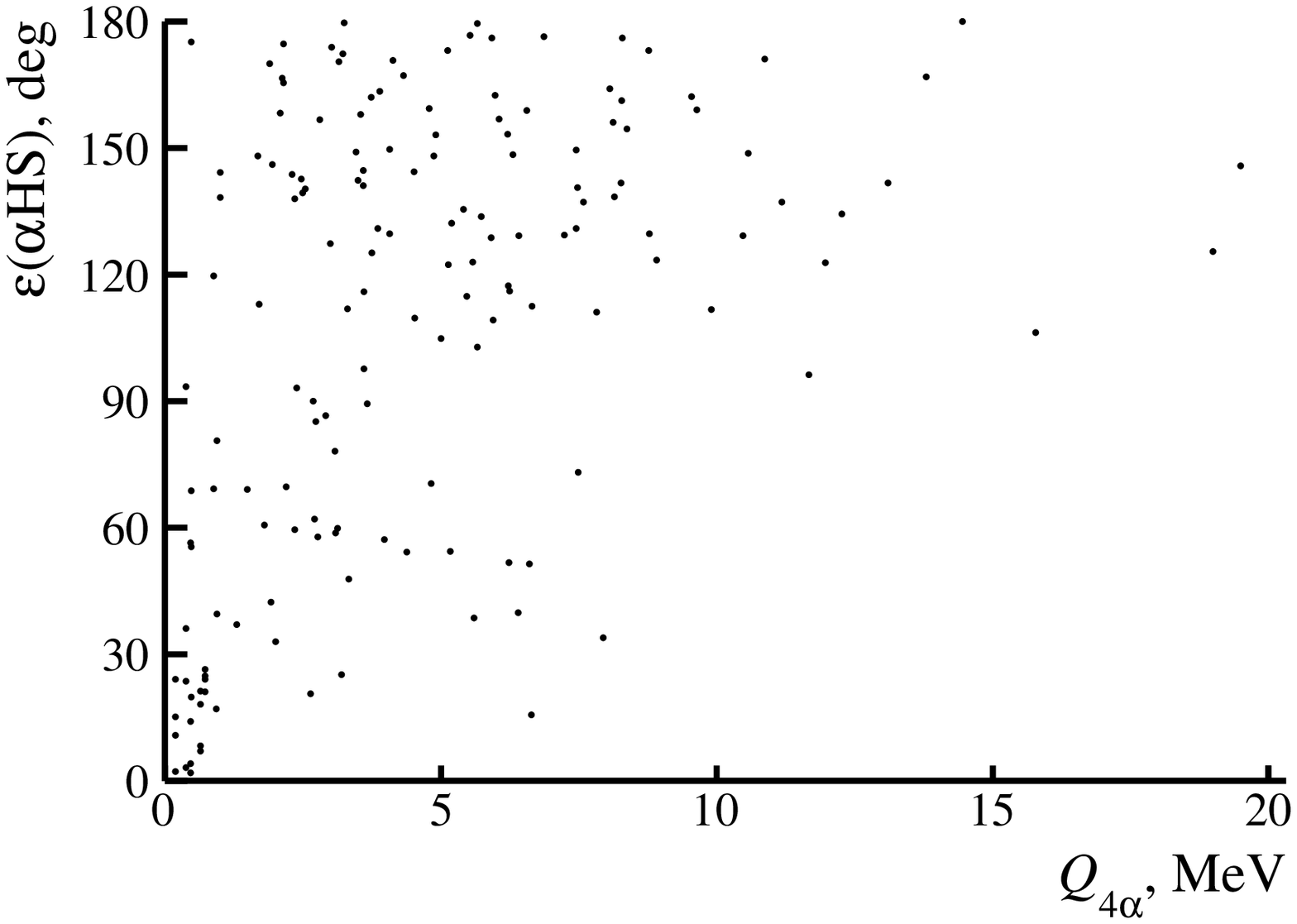}}
	\caption{Distribution of events ${}^{16}$O $\to$ $\alpha$HS over invariant mass $Q_{4\alpha}$ and azimuth angle $\varepsilon$($\alpha$HS).}
\end{figure}

HS can arise as a product of the $\alpha$ decay of the excited state 0$^+_6$ of the ${}^{16}$O nucleus \cite{5,6} (by analogy with the decay of HS into ${}^{8}$Be + $\alpha$). In the 641 ``white'' star, the full distribution of $\alpha$-quartets over $Q_{4\alpha}$ (Fig. 6) is described mainly by the Rayleigh distribution with the parameter $\sigma_{(Q4\alpha)}$ = (6.1 $\pm$ 0.2) MeV. The condition for the presence in the 4$\alpha$-event ($\alpha$HS) of at least one $\alpha$-triple with $Q_{3\alpha}$ $<$ 700 keV changes toward the low-energy direction the $Q_{4\alpha}$ distribution (Fig. 6) and the value of the parameter $\sigma_{Q4\alpha}$ = (4.5 $\alpha$ 0.5) MeV. It can be assumed that in the decay of an integral object, HS and $\alpha$ will be correlated in direction. Figure 7 shows the distribution of $\alpha$HS over $Q_{4\alpha}$ and the azimuth angle $\varepsilon$($\alpha$HS) between the HS directions and the $\alpha$ particle. It is worth noting that $Q_{4\alpha}$ and $\varepsilon$($\alpha$HS) are functionally related. The condition $\varepsilon$($\alpha$HS) $<$ 45$^\circ$ identifies 9 events that satisfy $Q_{4\alpha}$ $<$ 1 MeV with an average value of  $\langle Q_{4\alpha} \rangle$ = (624 $\pm$ 84) keV with RMS 252 keV (Fig. 6). On their basis, the assessment of the contribution of the 0$^+_6$ state is 7 $\pm$ 2\%.
 
\begin{figure}
	\centerline{\includegraphics*[width=0.55\linewidth]{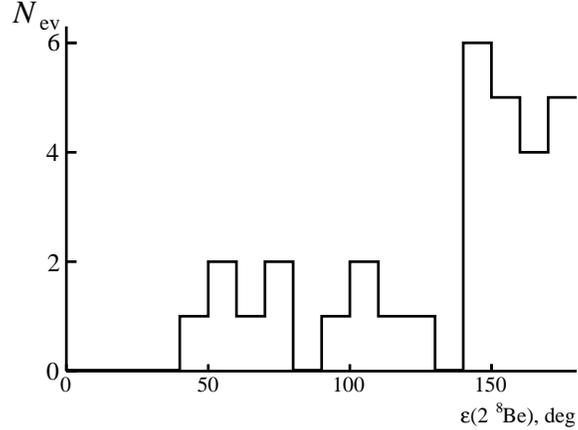}}
	\caption{Distribution of events ${}^{16}$O $\to$ 2${}^{8}$Be over azimuth angle $\varepsilon$(2${}^{8}$Be) between ${}^{8}$Be fragments.}
\end{figure}

\begin{figure}
	\centerline{\includegraphics*[width=0.55\linewidth]{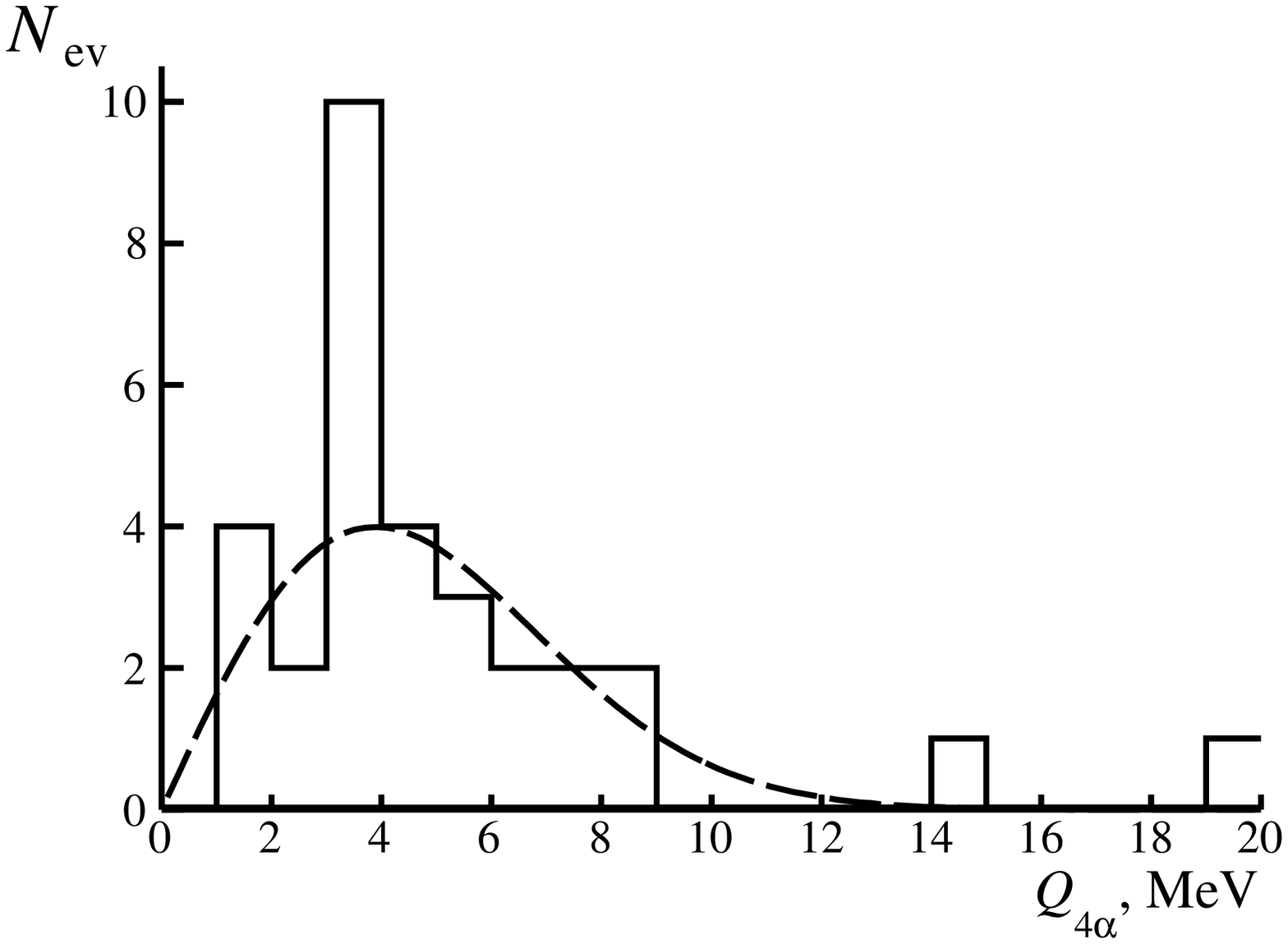}}
	\caption{Distribution of events ${}^{16}$O $\to$ 2${}^{8}$Be over invariant mass $Q_{4\alpha}$; line - Rayleigh distribution.}
\end{figure}

Among the ``white'' stars ${}^{16}$O $\to$ 4$\alpha$, 33 events were selected, in which two ${}^{8}$Be fragments ($Q_{2\alpha}$ $<$ 0.2 MeV) are present. The directions of expansion along the azimuth angle $\varepsilon$ (2${}^{8}$Be) show anti-correlation (Fig. 8) which indicates the binary formation of these fragments. In 31 events 2${}^{8}$Be there are no triples of $\alpha$-particles that satisfy the condition HS ($Q_{3\alpha}$ $<$ 700 keV) which gives an estimate of the contribution of the ${}^{16}$O $\to$ 2${}^{8}$Be channel is equal to 5 $\pm$ 1\%. The distribution of pairs to their full transverse momentum $P$T(2${}^{8}$Be) is described by the Rayleigh distribution with the parameter $\sigma$PT (2${}^{8}$Be) = (161 $\pm$ 2) MeV/$c$. Figure 9 shows the distribution over $Q_{4\alpha}$ for 2${}^{8}$Be events for which the Rayleigh parameter is (4.3 $\pm$ 1.2) MeV. The number of events in the channels ${}^{16}$O $\to$ $\alpha$HS and ${}^{16}$O $\to$ 2${}^{8}$Be has a ratio 4.5 $\pm$ 0.4 which means that the first of them is clearly leading.

\begin{figure}
	\centerline{\includegraphics*[width=0.55\linewidth]{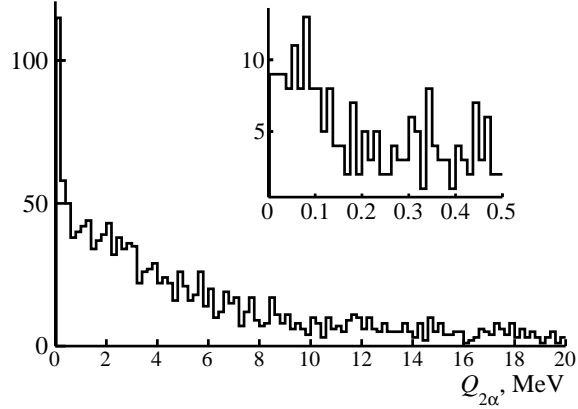}}
	\caption{Distribution over invariant mass $Q_{2\alpha}$ of all $\alpha$-pairs in the fragmentation channels ${}^{22}$Ne $\to$ (2-5)$\alpha$ at 3.22 $A$ GeV.}
\end{figure}

\section{FRAGMENTATION ${}^{22}$Ne}

The results of measurements of 4301 interaction of ${}^{22}$Ne nuclei at energy of 3.22 $A$ GeV are available for analysis. The search for events was performed by scanning tracks of primary nuclei (that is, without sampling) providing an overview of the ${}^{22}$Ne fragmentation topology \cite{9}. This set includes measurements of the angles of emission of relativistic $\alpha$-particles for 528 2$\alpha$, 243 3$\alpha$, 80 4$\alpha$, and 10 5$\alpha$ events which allows analysis in variables of the invariant mass $Q$(2-5)$\alpha$. It is worth noting that measurements of the angles of scattering of the fragments were made by the method that gives worse relative accuracy than in the cases presented above. Nevertheless, the $Q_{2\alpha}$ distribution allows one to isolate the ${}^{8}$Be signal in the region $Q_{2\alpha}$ $<$ 0.2 MeV (Fig. 10).

\begin{figure}
	\centerline{\includegraphics*[width=0.6\linewidth]{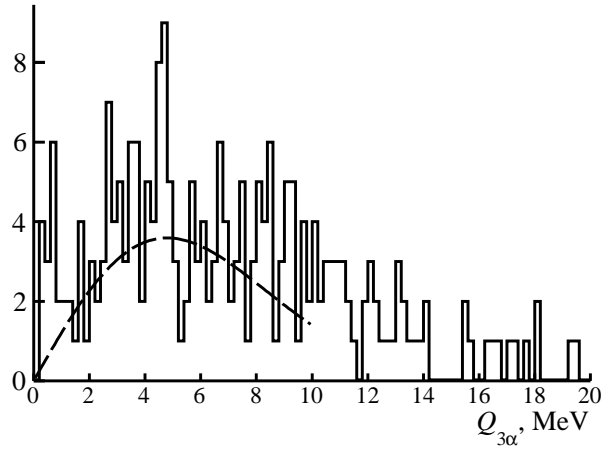}}
	\caption{Distribution over invariant mass $Q_{3\alpha}$ of all $\alpha$-triples in 4$\alpha$ channel in fragmentation of ${}^{22}$Ne nuclei at 3.22 $A$ GeV; line - Rayleigh distribution.}
\end{figure}

Figure 11 shows the distribution over $Q_{3\alpha}$ for channel 4$\alpha$. In the region $Q_{3\alpha}$ $<$ 1 MeV the average value is $\langle Q_{3\alpha} \rangle$ = (557 $\pm$ 51) keV and RMS 195 keV which is close to the HS value. The values of the Rayleigh distribution parameters $\sigma_{Q3\alpha}$ in approximations of $Q_{3\alpha}$ distributions for channels 3$\alpha$, 4$\alpha$ and 5$\alpha$ in the region up to 10 MeV are (4.0 $\pm$ 0.5), (4.3 $\pm$ 0.6) and (4.3 $\pm$ 0.4) MeV, respectively. These values are close to the cases of ${}^{12}$C and ${}^{16}$O. In the $Q_{3\alpha}$ $<$ 1 MeV region, the number of candidates for HS decays (and their share) in the 3$\alpha$, 4$\alpha$ and 5$\alpha$ channels is 3 (1.2 $\pm$ 0.7\%), 12 (15 $\pm$ 4\%) and 1 (10\%). Thus, only in the 4$\alpha$-channel there is a significant indication of HS. The shift $\langle Q_{3\alpha} \rangle$ compared with cases of ${}^{12}$C and ${}^{16}$O requires better measurement accuracy.

\begin{figure}
	\centerline{\includegraphics*[width=0.6\linewidth]{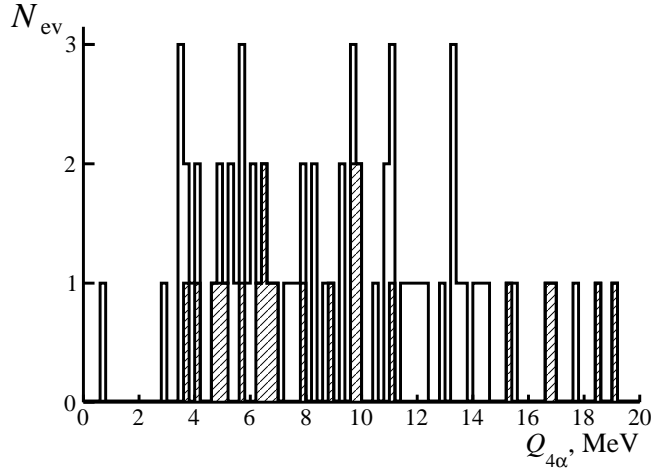}}
	\caption{Distribution of all $\alpha$-quartets in 4$\alpha$ and 5$\alpha$ channels over invariant mass $Q_{4\alpha}$ in the fragmentation of ${}^{22}$Ne nuclei at 3.22 $A$ GeV 5$\alpha$ (marked with hatching).}
\end{figure}

The distributions over $Q_{4\alpha}$ in channels 4$\alpha$ and 5$\alpha$ are presented in Figure 12. Its main part limited by $Q_{3\alpha}$ $<$ 10 MeV is described by the Rayleigh distribution with the parameter $\sigma_{Q4\alpha}$ = (4.9 $\pm$ 0.9) MeV. In the 4$\alpha$ channel there is a single event with the value $Q_{4\alpha}$ = 791 keV in which all $\alpha$-triples meet the condition HS $Q_{3\alpha}$ $<$ 1 MeV. This $\alpha$-quartet can correspond to the decay of the ${}^{12}$O 0$^+_6$ state. Obvious interest is the increase in statistics in channels 4$\alpha$ and 5$\alpha$. An additional possibility is provided by the existing NTE layers irradiated with ${}^{28}$Si nuclei at 3.65 $A$ GeV.

\section*{CONCLUSION}

Based on the data obtained in the 1980s – 1990s on dissociation of relativistic ${}^{12}$C, ${}^{16}$O and ${}^{22}$Ne nuclei in the nuclear track emulsion, as well as of their modern complement in the case of ${}^{12}$C a search for triples of relativistic $\alpha$-particles in the Hoyle state was performed. Determining the invariant mass of the $\alpha$-particle triples by their emission angles in the approximation of preserving the velocity of the parent nucleus ensures sufficient accuracy in identifying the HS against the background of higher 3$\alpha$ excitations of the ${}^{12}$C nucleus. The contribution of HS decays to ${}^{12}$C $\to$ 3$\alpha$ dissociation is 11 $\pm$ 3\%.

In the dissociation of ${}^{16}$O $\to$ 4$\alpha$, the contribution of HS decays is 22 $\pm$ 2\%. Attention is drawn to the fact that an increase in combinations of $\alpha$-particles leads to a noticeable increase in the contribution of HS to the dissociation of ${}^{16}$O $\to$ 4$\alpha$. An analysis of the invariant masses of $\alpha$-quartets gives an estimate of the contribution of the decays of the state ${}^{16}$O 0$^+_6$ to 7 $\pm$ 2\%. Consequently, the direct dissociation $\alpha$ + HS dominates in the HS formation.

Analysis of fragmentation of the ${}^{22}$Ne nucleus revealed the HS formation only in the 4$\alpha$ channel for which the share of events with HS was 15 $\pm$ 4\%. Having insufficient statistical security this result serves as a guideline for continuing the search for $\alpha$-ensembles by accelerating scanning over the area of nuclear NTE layers.

In general, the HS feature as a universal and sufficiently long-lived object similar to the unstable ${}^{8}$Be nucleus is confirmed. The closest source for verifying the HS universality is peripheral dissociation of the ${}^{14}$N nucleus in which the 3He + H channel leads, with a contribution of ${}^{8}$Be decays of about 25\% \cite{10}. Analysis of the NTE layers irradiated in the early 2000s with relativistic ${}^{14}$N nuclei was resumed in the context of the HS problem. A similar analysis will be carried out in the NTE layers which were irradiated by relativistic nuclei ${}^{22}$Ne and ${}^{28}$Si at the JINR Synchrophasotron in the late 80s and used for overview analysis. Despite the past decades this experimental material has retained the necessary quality.

\section*{ACKNOWLEDGMENTS}

The presented material is based on the analysis of experimental material and data obtained since the beginning of the 70s and up to the present. In itself, this fact demonstrates the solidity of the method of the nuclear track emulsion and its ability to evolve. It is impossible to list all the participants in the emulsion cooperation at the JINR Synchrophasotron in analyzing the interactions of relativistic nuclei. Behind this entire scientific heritage there was enormous work and the joy of being pioneers in. Therefore, we hope that citing publications and the further use of materials and methods are not only useful in research but also serve to preserve the memory of the era of the emergence of research on relativistic nuclear physics.
The authors are especially grateful to A.I. Lvov and E.P. Cherenkova for the opportunity to present the results of the research at the Cherenkov Readings at FIAN. Held annually, this event was in 2019 already the 12th in a row.

%
%


\begin{thebibliography}{10}
%

\bibitem {1}
Zarubin P.I. // Lect. Notes in Physics 2014, V. 875, \href{https://link.springer.com/chapter/10.1007/978-3-319-01077-9_3}{Clusters in Nuclei}, 3, P. 51, Springer Int. Publ.; \href{https://arxiv.org/abs/1309.4881}{arXiv:1309.4881}. 

\bibitem {2}
Artemenkov D.A., Zaitsev A.A., Zarubin P.I. // Phys. Part. Nucl. 2017, V. 48, P. 147; DOI:  \href{https://link.springer.com/article/10.1134%2FS1063779617010026}{10.1134/S1063779617010026} ; \href{https://arxiv.org/abs/1607.08020}{arXiv:1607.08020}.

\bibitem {3}
Artemenkov D.A., Bradnova V., Britvich G.I., Firu E., Haiduc M., Kalinin V.A., Kharlamov S.P., Kornegrutsa N.K., Kostin M.Yu., Maksimov A.V., Mitseva E., Neagu A., Pikalov V.A., Polkovnikov M.K., Rusakova V.V., Stanoeva R., Zaitseva, A.A., Zarubin P.I., Zarubina I.G. // Rad. Meas. 2018, V. 119, P. 199;  DOI: \href{https://www.sciencedirect.com/science/article/pii/S1350448718301094?via%3Dihub} {10.1016/j.radmeas.2018.11.005};  \href{https://arxiv.org/abs/1812.09096}{arXiv:1812.09096}.

\bibitem {4}
Freer M., Fynbo H.O.U. // Prog, in Part. and Nucl. Phys. 2014, V. 78, P. 1; DOI: \href{https://www.sciencedirect.com/science/article/pii/S0146641014000453} {10.1016/j.ppnp.2014.06.001}.

\bibitem {5}
Tohsaki A., Horiuchi H., Schuck P. and Ropke G. // Rev. Mod. Phys. 2017, V. 89, 011002; DOI: \href{https://journals.aps.org/rmp/abstract/10.1103/RevModPhys.89.011002}{10.1103/RevModPhys.89.011002}.

\bibitem {6}
Schuck P. // \href{https://arxiv.org/abs/1811.11580}{arXiv:1811.11580}.

\bibitem {7}
Belaga V.V., Benjaza A.A., Rusakova V.V., Salomov D.A., Chernov G.M., Phys. Atom. Nucl. 1995, V. 58, P. 1905; DOI: 10.1063/7788-1905(95)5811-5; \href{https://arxiv.org/abs/1109.0817}{arXiv:1109.0817}.

\bibitem {8}
Andreeva N.P. et al. Phys. Atom. Nucl. 1996, V. 59, P. 102; DOI: 10.1063/S5901-0102(96)7788-0; \href{https://arxiv.org/abs/1109.3007}{arXiv:1109.3007}.

\bibitem {9}
El-Naghy A. et al. J. Phys. G: Nucl. Phys. 1988, V. 14, P. 1125; DOI: \href{https://iopscience.iop.org/article/10.1088/0305-4616/14/8/015}{10.1088/0305-4616/14/8/015}.

\bibitem {10}
Shchedrina T.V. et al. Phys. Atom. Nucl. 2007, V. 70, P. 1230; DOI: \href{https://link.springer.com/article/10.1134%2FS1063778807070149}{10.1134/S1063778807070149}; \href{https://arxiv.org/abs/nucl-ex/0605022}{arXiv:nucl-ex/0605022}.

\end{thebibliography}
\end{document}